# A proposal for Marchenko-based target-oriented full waveform inversion

S. M. Aydin Shoja, Giovanni A. Meles and Kees Wapenaar

## Abstract

The Hessian matrix plays an important role in correct interpretation of the multiple scattered wave fields inside the FWI frame work. Due to the high computational costs, the computation of the Hessian matrix is not feasible. Consequently, FWI produces overburden related artifacts inside the target zone model, due to the lack of the exact Hessian matrix. We have shown here that Marchenko-based target-oriented Full Waveform Inversion can compensate the need of Hessian matrix inversion by reducing the non-linearity due to overburden effects. This is achieved by exploiting Marchenko-based target replacement to remove the overburden response and its interactions with the target zone from residuals and inserting the response of the updated target zone into the response of the entire medium. We have also shown that this method is more robust with respect to prior information than the standard gradient FWI. Similarly to standard Marchenko imaging, the proposed method only requires knowledge of the direct arrival time from a focusing point to the surface and the reflection response of the medium.

## Introduction

Nowadays, the interest for inversion of a relatively small target zone of the subsurface, especially, for reservoir monitoring applications, is increasing. One of the main inversion techniques for this purpose is full waveform inversion (FWI). The Hessian matrix plays an important role in correct interpretation of the multiple scattered wave fields inside the FWI frame work (Metivier, et al., 2017). Due to the high computational costs, the computation of the Hessian matrix is not feasible. Consequently, FWI produces overburden related artifacts inside the target zone model, due to the lack of the exact Hessian matrix.

Different target-oriented approaches have been proposed to compensate for the lack of the exact Hessian: Data redatuming techniques (Yang, et al., 2012) and model domain cost functions (Tang, 2009), to name but a few.

Recently, Marchenko-based target replacement has been introduced as a method to predict the response of the overburden and remove the response of the target zone and insert the response of a new one into the response of the medium. This method only needs a smooth model of the overburden and a surface reflection data (Wapenaar & Staring, 2018). With this, one can do target-oriented FWI without the need for data redatuming or computing model domain cost functions.

First, a short description of full waveform inversion is given followed by a short explanation of Marchenko-based target replacement. Next, we combine these methods to obtain Marchenko-based target-oriented FWI. Finally, this method is validated through a numerical test.

## Full waveform inversion

In general, full waveform inversion is formulated as a partial differential equation constrained optimization problem in which a data-driven cost function is minimized with the constraint of solving the wave equation (Metivier, et al., 2017). This cost function is defined as the square of the $L_2$ norm of the data residuals:

$$C(\mathbf{m}) = \left\| \mathbf{d}^{pred}(\mathbf{m}) - \mathbf{d}^{obs} \right\|_2^2. \quad (1)$$

Here, $\mathbf{d}^{pred}$ is the predicted data vector and $\mathbf{d}^{obs}$ is the observed data vector. In order to minimize this cost function, the gradient-based optimization methods are used. The gradient of this cost function with respect to the model parameters is (Virieux & Operto, 2009):

$$\nabla C_\mathbf{m} = -\Re \left[ \mathbf{J}^\dagger \Delta \mathbf{d} \right]. \quad (2)$$

Here, the $\dagger$ symbol is complex conjugate transpose, $\mathbf{J}$ is the Fréchet derivative matrix, $\Delta \mathbf{d}$ is the data residuals vector and $\Re$ denotes the real part. It is possible to find an expression for the Fréchet derivative matrix in terms of Green's functions by considering the Born approximation and taking the slowness $s(\mathbf{x})$ as the model parameter (Schuster, 2017):

$$\frac{\partial \mathrm{d}^{pred}(\mathbf{x}_r, \mathbf{x}_s, \omega)}{\partial s(\mathbf{x})} = 2\omega^2 s(\mathbf{x}) G(\mathbf{x}_r, \mathbf{x}, \omega) W(\omega) G(\mathbf{x}, \mathbf{x}_s, \omega), \quad (3)$$

where $G(\mathbf{x}, \mathbf{x}_s)$ and $G(\mathbf{x}_r, \mathbf{x})$ are Green's functions from source to the scatterer and from scatterer to the receiver location respectively and $W(\omega)$ is the source wavelet. Therefore, the gradient at the position of a model parameter can be rewritten as:

$$\nabla C_\mathbf{m}(\mathbf{x}) = -2\Re \left[ s(\mathbf{x}) \int \omega^2 G(\mathbf{x}_r, \mathbf{x}) W(\omega) G(\mathbf{x}, \mathbf{x}_s) \Delta \mathbf{d}^* d\omega \right]. \quad (4)$$

## Marchenko-based target replacement

Wapenaar and Staring (2018) devised a method to remove the effects of a target zone inside the medium from the reflection response of the entire medium and insert a changed target zone inside the medium response.

To this end, they employed the one-way reciprocity theorems to derive a representation for the reflection response of the entire medium in terms of responses of the overburden and the target zone:

$$R_B^{\cup}(\mathbf{x}_r, \mathbf{x}_s, \omega) = R_A^{\cup}(\mathbf{x}_r, \mathbf{x}_s, \omega) + \int_{S_1}\int_{S_1} T_A^{-}(\mathbf{x}_r, \mathbf{x}', \omega) R_b^{\cup}(\mathbf{x}', \mathbf{x}, \omega) G_B^{+,+}(\mathbf{x}, \mathbf{x}_s, \omega) d\mathbf{x} d\mathbf{x}'. \qquad (5)$$

Here, A, B, and b refer to overburden, entire medium and target zone respectively. $R^{\cup}$ is the reflection response of that medium from above, $T^{-}$ is the upward propagating transmission response and $G^{+,+}$ is the downward propagating Green's function of a downward emitting source. $\mathbf{x}_s$ and $\mathbf{x}_r$ are located just above the surface ($S_0$). $\mathbf{x}$ and $\mathbf{x}'$ are located at $S_1$, a transparent boundary between the overburden and the target zone.

In order to find the responses of overburden, i.e. medium A, one can apply the Marchenko method to reflection response $R_B^{\cup}$ to find the so-called focusing functions and use Multi-dimensional deconvolution (MDD) to resolve the responses of the medium A. The Green's function $G_B^{+,+}(\mathbf{x}, \mathbf{x}_s, \omega)$ can be retrieved by inverting the following relations:

$$T_A^{+}(\mathbf{x}'', \mathbf{x}_s, \omega) = \int_{S_1} C_{Ab}(\mathbf{x}'', \mathbf{x}, \omega) G_B^{+,+}(\mathbf{x}, \mathbf{x}_s, \omega) d\mathbf{x}, \qquad (6)$$

where

$$C_{Ab}(\mathbf{x}'', \mathbf{x}, \omega) = \delta(\mathbf{x}''_H - \mathbf{x}_H) - \int_{S_1} R_A^{\cap}(\mathbf{x}'', \mathbf{x}', \omega) R_b^{\cup}(\mathbf{x}', \mathbf{x}, \omega) d\mathbf{x}'. \qquad (7)$$

**Marchenko-based target-oriented Full Waveform Inversion**

In each iteration of FWI a new updated model of the target zone is generated. Using equation (5) and doing modeling just inside the target zone, it is possible to insert the response of the updated target zone into the reflection response of the medium and use it as the new predicted data. Let's explain it in more detail. Consider $b_i$ as the model parameter of the target zone in each iteration and denote changed quantities with an overbar. By modeling inside the target zone in each iteration the reflection response of $b_i$, $\overline{R}_{b_i}^{\cup}(\mathbf{x}', \mathbf{x}, \omega)$, is generated. Then, by applying the Marchenko method and Multi-dimensional deconvolution and inverting equations (6) and (7) one can calculate $\overline{G}_{B_i}^{+,+}(\mathbf{x}, \mathbf{x}_s, \omega)$, $T_A^{-}(\mathbf{x}_r, \mathbf{x}', \omega)$ and $R_A^{\cup}(\mathbf{x}_r, \mathbf{x}_s, \omega)$ with $\mathbf{x}$ and $\mathbf{x}'$ at $S_1$, and $\mathbf{x}_s$ and $\mathbf{x}_r$ at $S_0$. Thus:

$$\mathbf{d}_i^{pred} = \overline{R}_{B_i}^{\cup}(\mathbf{x}_r, \mathbf{x}_s, \omega), \qquad (8)$$

and

$$\Delta \mathbf{d}_i = \overline{R}_{B_i}^{\cup}(\mathbf{x}_r, \mathbf{x}_s, \omega) - \mathbf{d}^{obs}. \qquad (9)$$

Since the first term of the equation (5) is the response of the overburden and it also exists inside the observed data, by computing the data residuals the response of the overburden is completely removed and the data residuals only contain the response of the target zone in each iteration.

Until now, a method has been presented for making the predicted data without knowing the overburden model and removing the effects of the overburden from the data residuals. Since the gradient of the cost function needs the Green's functions inside the target zone with a source at $S_0$ (see equation(4)), for the next step these Green's functions need to be calculated.

Let's call the Green's functions inside the target zone with $\mathbf{x}'$ at $S_1$ and $\mathbf{x}$ variable inside $b_i$, $\overline{G}_{b_i}^{p,+}(\mathbf{x}, \mathbf{x}', \omega)$, and the Green's functions inside the target zone with $\mathbf{x}_s$ at $S_0$, $\overline{G}_{B_i}^{p,+}(\mathbf{x}, \mathbf{x}_s, \omega)$, where superscript p means the whole Green's function, i.e. $G^{p,+} = G^{-,+} + G^{+,+}$. By inverting equations (6) and

(7) it is possible to find $\overline{G}_{B_i}^{+,+}(\mathbf{x}'',\mathbf{x}_s,\omega)$ where $\mathbf{x}''$ is at $S_1$. Finally, by using $\overline{G}_{b_i}^{p,+}(\mathbf{x},\mathbf{x}',\omega)$ as a propagator one can make $\overline{G}_{B_i}^{p,+}(\mathbf{x},\mathbf{x}_s,\omega)$:

$$\overline{G}_{B_i}^{p,+}(\mathbf{x},\mathbf{x}_s,\omega) = \int_{S_1} \overline{G}_{b_i}^{p,+}(\mathbf{x},\mathbf{x}',\omega)\overline{G}_{B_i}^{+,+}(\mathbf{x}',\mathbf{x}_s,\omega)d\mathbf{x}'. \qquad (10)$$

Considering the Green's functions reciprocity, this Green's function, i.e. $\overline{G}_{B_i}^{p,+}(\mathbf{x},\mathbf{x}_s,\omega)$, is used for both Green's functions inside the equation (4).

**Numerical examples**

In order to confirm the effectiveness of this method, a comparison between FWI for the entire medium and Marchenko-based target-oriented FWI with a one-dimensional acoustic model with a constant density (Figure 1) was done. For this purpose, a gradient descent algorithm is used. A delta function with a time sampling of $10^{-2}$ seconds is used as the source signature and the depth sampling is set to 10 meters. For the Target-oriented case, a focusing depth of 2800 meters is chosen. In figure (2) a comparison between the retrieved velocity models is shown and in figure (4) cost functions are illustrated. In addition, in figure (3) residual vectors are compared.

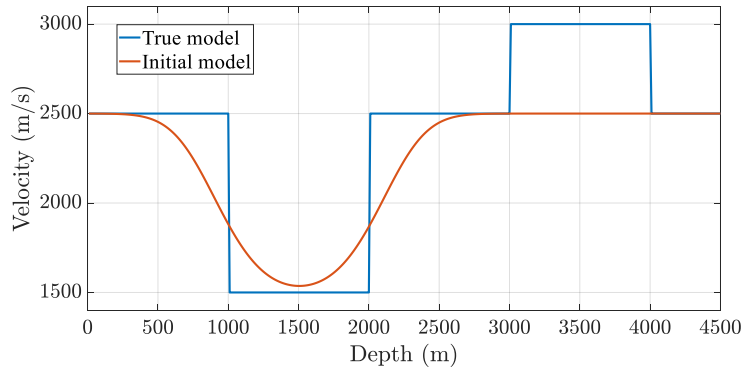

**Figure 1** True and initial model

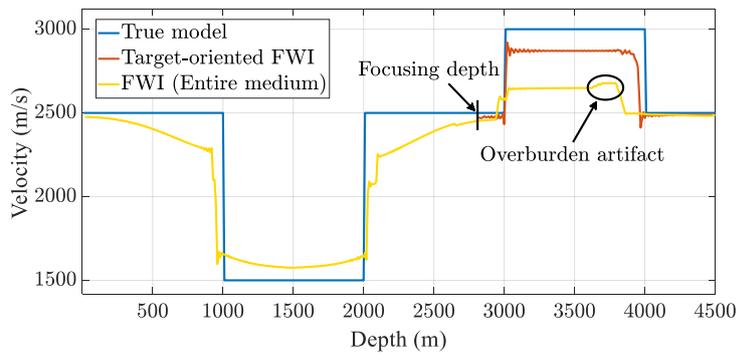

**Figure 2** Comparison between FWI for the entire medium and Marchenko-based target-oriented FWI. The internal multiple of the overburden created an artifact inside the target zone, but it is disappeared from the results of the Target-oriented FWI.

These results clearly show that this proposed method is able to remove the overburden multiple reflection artifacts from the updated model of the target zone. It also produced a more accurate model of the target zone in terms of amplitude and reflector positioning.

**Conclusion**

We have shown here that Marchenko-based target-oriented Full Waveform Inversion can compensate the need of Hessian matrix inversion by reducing the non-linearity due to overburden effects. This is achieved by exploiting Marchenko-based target replacement to remove the overburden response and its interactions with the target zone from residuals (see figure (3)) and inserting the response of the updated target zone into the response of the entire medium. With a 1D model we have also shown that this method is more robust with respect to prior information than the standard gradient FWI. Similarly to standard Marchenko imaging, the proposed method only requires knowledge of the direct arrival time from a focusing point to the surface and the reflection response of the medium.

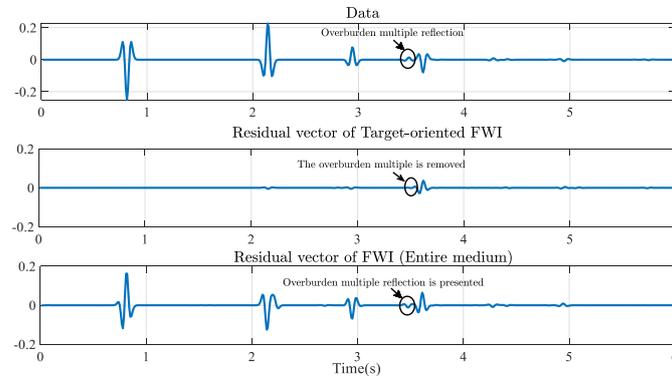

**Figure 3** Comparison between Observed data, the residual vector of Target-oriented FWI and FWI for the entire medium. The overburden response is removed from the residuals of the target-oriented FWI by the Marchenko-based target replacement method. Whereas, it is still presented in FWI for the entire medium. For better visualization, traces are convolved with a Ricker wavelet with a dominant frequency of 40 Hz.

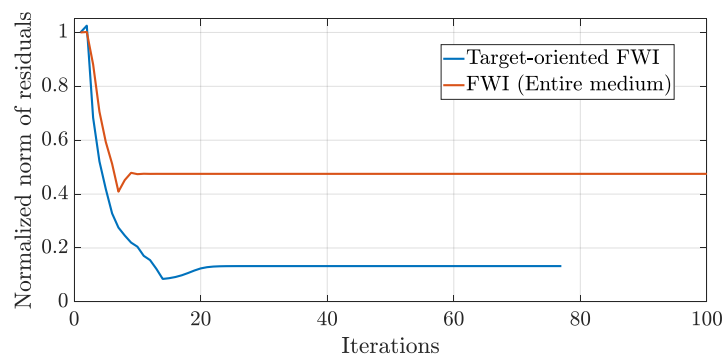

**Figure 4** Comparison between cost functions

**Acknowledgment**

This research was funded by the European Research Council (ERC) under the European Union's Horizon 2020 research and innovation programme (grant agreement No: 742703).